# Theory, observation, and ultrafast response of novel hybrid anapole states

**Adrià Canós Valero[1], Egor A. Gurvitz[1], Fedor A. Benimetskiy[1], Dmitry A. Pidgayko[1], Anton Samusev[1], Andrey B. Evlyukhin[1,2], Dmitrii Redka[3], Michael.I.Tribelsky[4,5,6], Mohsen Rahmani[7], Khosro Zangeneh Kamali[8], Alexander A. Pavlov[9], Andrey E. Miroshnichenko[10] and Alexander S. Shalin[1,11]**

[1]ITMO University, Kronverksky prospect 49, 197101, St. Petersburg, Russia

[2]Moscow Institute of Physics and Technology, 9 Institutsky Lane, Dolgoprudny 141700, Russia

[3]Electrotechnical University "LETI" (ETU), 5 Prof. Popova Street, Saint Petersburg 197376, Russia

[4]Faculty of Physics, M. V. Lomonosov Moscow State University, Moscow 119991, Russia

[5]National Research Nuclear University, Moscow Engineering Physics Institute, Moscow 115409, Russia

[6]RITS Yamaguchi University, Yamaguchi 753-8511, Japan

[7]Advanced Optics and Photonics Laboratory, Department of Engineering, School of Science and Technology, Nottingham Trent University, Nottingham, NG11 8NS, UK

[8]1ARC Centre of Excellence for Transformative Meta-Optical Systems, Research School of Physics, The Australian National University, Canberra ACT 2601, Australia

[9]Institute of Nanotechnology of Microelectronics of the Russian Academy of Sciences (INME RAS), Moscow, Nagatinskaya street, house 16A, building 11

[10]School of Engineering and Information Technology, UNSW Canberra, ACT, 2600, Australia

[11]Ulyanovsk State University, Lev Tolstoy Street 42, 432017, Ulyanovsk, Russia

**Abstract.** Modern nanophotonics has witnessed the rise of "electric anapoles", destructive interferences of electric dipoles and toroidal electric dipoles, actively exploited to cancel electric dipole radiation from nanoresonators. However, the inherent duality of Maxwell's equations suggests the intriguing possibility of "magnetic anapoles", involving a nonradiating composition of a magnetic dipole and a magnetic toroidal dipole. Here, we predict, fabricate and observe experimentally via a series of dark field spectroscopy measurements a hybrid anapole of mixed electric and magnetic character, with all the dominant multipoles being suppressed by the toroidal terms in a nanocylinder. We delve into the physics of such exotic current configurations in the stationary and transient regimes and predict a number of ultrafast phenomena taking place within sub-ps times after the breakdown of the hybrid anapole. Based on the preceding theory, we design a non-Huygens metasurface featuring a dual functionality: perfect transparency in the stationary regime and controllable ultrashort pulse beatings in the transient.

## Introduction

Over the past few years, all-dielectric nanophotonics has become one of the cornerstones of modern research in nano-optics[1]. Unlike plasmonic structures, dielectric ones allow overcoming the fundamental limitation of Ohmic losses. Utilizing electric and magnetic Mie-like resonances of nanoparticles consisting of low-loss high-index semiconductor or dielectric materials, such as Si, TiO$_2$, Ge, GaAs[2,3], enables manipulating both the electric and magnetic components of light at the nanoscale. This emerging field has already led to a wide range of exciting applications, such as low-loss discrete dielectric waveguides[4,5], passive and reconfigurable directional sources[6,7], efficient high harmonic generation mechanisms[8], light-harvesting and antireflective coatings[9–11], all-dielectric metasurfaces with artificially tailored optical response[12–15], dielectric beam deflectors[16], subwavelength all-optical liquid mixing[17], to mention just a few.

The ability to properly describe and predict electromagnetic scattering is of prime importance to manipulate the behavior of light at the nanoscale. For this purpose, different types of electromagnetic multipole expansions were introduced[18–22]. Among them, the charge-current Cartesian decomposition is widely used for describing optical signatures of nano-objects of arbitrary shape[18–20]. One of the most intriguing possibilities delivered by this formalism is the ability to define the so-called toroidal moment family[23–26]. While the well-known electric toroidal dipole moment is associated with the poloidal currents flowing along the meridians of a torus [24], higher-order toroidal moments, also known as mean square radii, feature more complex current distributions recently investigated theoretically in Refs.[27,28].

The electric toroidal dipole is now widely exploited in nanophotonics and metamaterials, active photonics[29], ultrasensitive biosensing[23] and applications requiring strong near field localization[30,31]. The fields radiated by toroidal moments share the same angular momentum and far-field properties as their electric or magnetic multipolar counterparts, allowing for the realization of two exciting effects: (i) enhanced multipolar response[32] enabled by the constructive interference of the fields, and (ii) mutual cancellation of the far-field contributions via the destructive one, so-called "anapole" states[33].

In the aforementioned scenario, an anapole state corresponds to a scattering minima from a given multipole channel[34,33], leading to a reduction of the overall far-field scattering and confined near fields[33]. This promising feature has motivated a widespread of investigations in diverse applications of nanoscale light-matter interactions ranging from photocatalysis[35], Raman scattering[36], strong exciton coupling[37] and second and third harmonic generation[38–41]. While there exists a large amount of literature regarding the stationary (frequency domain) response of such states, emergent studies point out that a completely different picture may take place in the transient (non-harmonic time dependent) regime[42]. In this regard, the understanding and control of the transient response of anapoles is yet to be developed, and remains an unexplored realm with potential applications in the novel field of ultrafast dynamic resonant phenomena[43,44].

Furthermore, the vast majority of the investigations on anapole states are limited to the electric dipole term only[45–48]. No experimental works have proven the existence of magnetic anapole states. Were the anapoles of electric and magnetic nature to spectrally overlap, it would lead to unprecedently confined near fields and provide new exciting degrees of freedom for the design of light-matter interactions. Until now, this has remained a challenging task, since the formation of the so-called 'hybrid anapole states' requires a careful overlap of electric and magnetic multipoles with their toroidal counterparts. Indeed, magnetic toroidal moments themselves have only been observed in a very recent work, where a complex cluster of nanodisks was necessary to induce the desired multipole response[49].

Spherically symmetric structures are unable to support hybrid anapole states[50], since the latter always remain hidden by the contributions of other multipole moments with non-negligible scattering. Nevertheless, recent developments in the theory of multipole expansions[19,28] have now opened the possibility to qualitatively and quantitatively investigate higher order electric and magnetic anapole states in scatterers with arbitrary shape, beyond the limitations of the quasistatic regime.

Here we design and demonstrate experimentally for the first time the existence of such exotic states, schematically illustrated in **Figure 1a**. In the stationary regime, we theoretically predict a novel type of non-trivial, non-scattering state, accompanied by an effective internal field concentration, governed by the exceptional spectral overlap of four electric and magnetic anapole states of different orders. This pure *Hybrid Anapole* state originates from the far-field destructive interference of *all* the leading electric and magnetic Cartesian multipoles of a finite cylindrical scatterer with their associated toroidal moments. By doing so, we also prove for the first time that higher order (quadrupolar) toroidal moments can be excited in an isolated dielectric nanoparticle and contribute to essential features of its scattering response. Complementing the multipolar approach, the near-field maps are interpreted in terms of the eigenstates of an open cavity[51,52] (quasi-normal modes, QNMs), thus providing a complete physical picture of the effect and establishing a general theoretical framework valid for anapoles of arbitrary order, in scatterers of arbitrary shape.

We fabricate a series of individual silicon nanocylinders supporting the effect and validate experimentally our theoretical predictions via dark field spectroscopy measurements. Contrarily to a conventional anapole state, the hybrid anapole preserves its nonradiating nature in the presence of any dielectric substrate. We unveil the physical mechanism behind such a counterintuitive effect, and demonstrate its breakdown in the transient regime, where the spatiotemporal maps reveal relevant changes in the beating pattern of the decaying modes that explicitly depend on the substrate refractive index. Finally, based on the preceeding theory, we design and demonstrate a dual-functional metasurface consisting of hybrid anapole particles displaying full transparency in the stationary regime, and highly tunable spatiotemporal response in the transient one.

**The Cartesian multipole expansion and high-order anapole conditions**

The analysis of the optical response of a nanoparticle is usually carried out via the decomposition of the scattering cross section as a sum of multipoles, which represent independent scattering channels of the object. Here we utilize the irreducible Cartesian multipole expansion derived in Ref.[28] (for completeness, also given in the Supporting Information), that explicitly takes into account higher order toroidal moments. The latter interpretation is our starting point towards the physical understanding of higher-order anapole states.

Within this approach, an electric or magnetic anapole of order *n* in a subwavelength scatterer is given by the condition:

$$P^{(e,m)}_{i_1...i_n} + i\frac{k_d}{v_d}T^{(e,m)}_{i_1...i_n} = 0 \qquad (1)$$

Here we have denoted *n*th order electric or magnetic moments with $P^{(e)}$ and $P^{(m)}$ and corresponding electric and magnetic toroidal moments with $T^{(e)}$ and $T^{(m)}$, respectively. The number of subscripts indicates the order of each Cartesian tensor, i.e., one subscript corresponds to dipole, two correspond to quadrupole, etc. $k_d, \varepsilon_d$ are the wavenumber and the dielectric permittivity of the host medium, and $v_d$ is the speed of light in the medium. A hybrid anapole state occurs when more than one multipole moment fulfils Eq.(1) at a given wavelength, resulting in a simultaneous suppression of scattering of two or more channels. However, as mentioned above, light, in general, can be radiated out through other non-zero multipole moments, destroying the overall effect. Thus, only the cancellation of all the leading multipoles can enable a true hybrid anapole.

For the sake of clarity, in the rest of the manuscript, we will rely on the widespread notation for low-order multipoles, i.e. $p$, $m$ for electric and magnetic dipoles, and $Q^{(e)}$, $Q^{(m)}$ for electric and magnetic quadrupoles.

## Results

### Observation and multipole analysis of hybrid anapole states

Under conventional plane wave illumination, hybrid anapole states of homogeneous spherical particles are hidden by the contributions of high order multipoles [50,53]. We will show that this restriction naturally vanishes for nano-objects with additional geometrical degrees of freedom, like finite cylinders or parallelepipeds. Throughout this work, and particularly in the next section, we will unveil the fundamental reason behind this unusual behavior.

Let us consider a cylindrical silicon nanoparticle in air. Starting now, we will use amorphous silicon (a-Si) in both theoretical and experimental studies (for details refer to the Supporting Information). The illumination scheme is presented in the left inset of **Figure 1**b (normally incident $x$-polarized plane wave propagating along $-z$ direction).

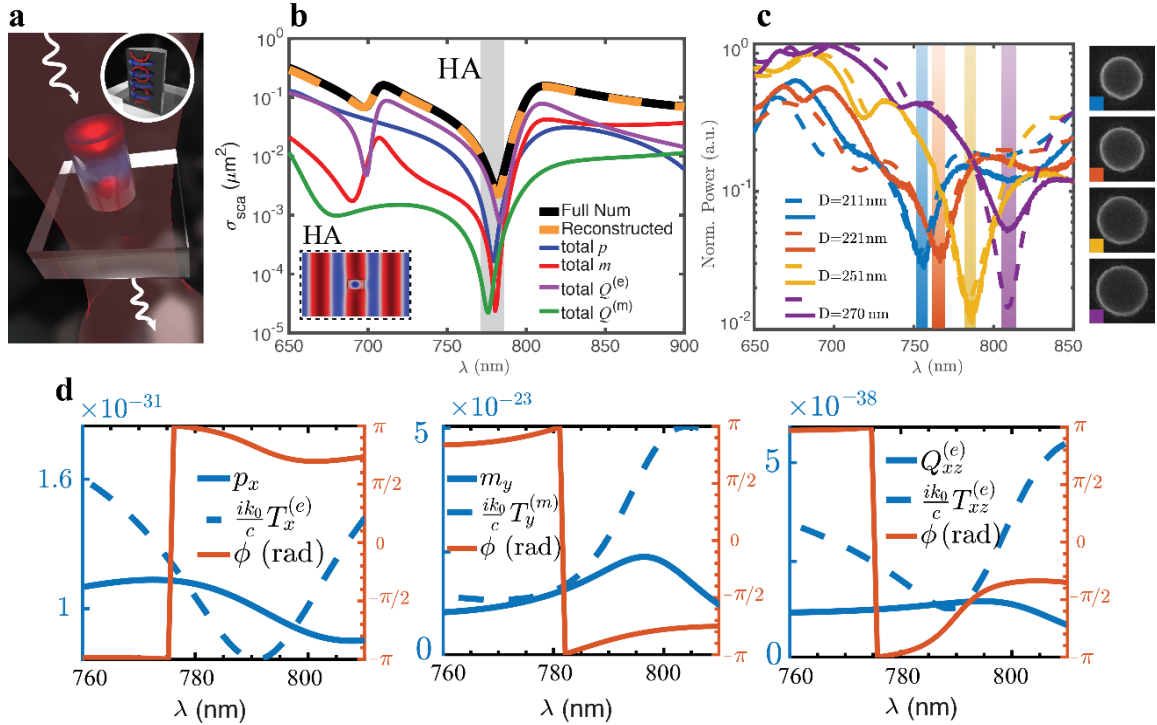

**Figure 1.** (a) Artistic representation of the novel effect. A normally incident plane wave excites nontrivial modal contributions in a Si nanocylinder whose interference with the background field leads to a four-fold hybrid anapole state yielding the nanoantenna virtually invisible in the far-field, with localized near field. Inset depicts the current distributions of the two resonant eigenmodes arising due to standing waves between the top and bottom walls (red) and lateral walls (blue) in the vertical plane. (b) Multipole reconstruction of the numerically obtained scattering cross section for the cylindrical amorphous silicon nanoparticle. In the legend caption, "total" implies that both basic and toroidal contributions of a given multipole are plotted. The inset corresponds to the x-component of the electric field. The geometrical parameters of the cylinder are height H=367 nm, diameter D=252 nm. Point HA ( $\lambda = 782$ nm ) corresponds to the hybrid anapole state. (c) Left plot: measured (solid lines) and simulated (dashed lines) scattering spectra of single isolated nanocylinders with different diameters D. The spectral positions of the hybrid anapoles are indicated by the colored regions. Right plot: SEM micrographs of the corresponding nanocylinder samples. The colored edges in each micrograph are associated to the legend entries in the measurements. (d) Amplitudes and phase differences between the multipoles and their toroidal counterparts. Panels from left to right, respectively: the basic electric and electric toroidal dipoles, the basic magnetic and magnetic toroidal dipoles, and the basic electric and electric toroidal quadrupoles. Amplitudes correspond to the left ordinate-axis, and phase differences are read from the right ordinate-axis.

The design methodology is based on the following: We note that the spectral positions of the full (basic and toroidal parts) electric dipole and magnetic quadrupole anapoles are mainly dependent on the cylinder's radius, while the wavelengths of the full magnetic dipole and electric quadrupole anapoles change both as functions of the cylinder height and radius. Figure S3 in the Supporting Information illustrates the spectral behavior of the multipolar anapoles with variations of the geometrical parameters in detail. Thus, carefully tuning these two geometrical degrees of freedom makes it possible to place the anapoles of all the leading terms ultimately close to each other (**Figure 1**b), providing a strong scattering minimum (**Figure 1**b-c, point HA).

The total scattering cross section and its multipole decomposition after the numerical optimization are shown in **Figure 1**b. Perfect agreement between the sum of the multipole contributions given by Eq.(1) in the Supporting Information and the result of the full-wave simulations in Comsol Multiphysics is demonstrated, proving that only the first four multipoles are sufficient to fully describe the optical response of the nanocylinder in the visible range. Therefore, the low-scattering regime delivered by this state renders it perfectly dark to the incident radiation (see the inset on **Figure 1**b).

The different panels in **Figure 1**d show the amplitudes and phase differences of the three most relevant multipoles with their toroidal moments. The results further confirm that the generalized anapole condition in Eq.(1) is well fulfilled for each pair (the amplitudes are equal, and they are $\pi$ rad out of phase) at the hybrid anapole wavelength $\lambda = 782$ nm. Particularly, this implies that we have succeeded in exciting, for the first time, toroidal quadrupole moments in the visible range.

To confirm our theoretical predictions on the novel hybrid anapole state, we have carried out direct scattering spectroscopy measurements for a set of standalone nanocylinders with tailored dimensions in the optical spectral range (**Figure 1**c). The measured scattering spectra (solid lines, **Figure 1**c) exhibit a pronounced dip, shifting with the increase of the nanocylinder diameter D, in excellent agreement with the calculations (dashed lines, **Figure 1**c). Technical details on the fabrication and the optical measurement setup can be found in the Supporting Information. While an increase of the lateral size of the nanocylinder leads to an overall redshift of the multipole anapoles, they almost perfectly overlap at $D = 251$ nm, where the most pronounced hybrid anapole state results in a large drop in scattering efficiency, of almost two orders of magnitude, rendering the nanocylinder virtually invisible. We note that the other dips indicated in the measured spectra also correspond to hybrid anapoles, although their overlap is not as much pronounced, but still results in a significant scattering reduction.

Noteworthy, the near-zero values of the full scattering coefficients do not imply the induced polarization currents in the particle to be also close to zero. This is in agreement with the usual anapole behavior[54], but in the new state, due to the suppression of several multipoles simultaneously, the hybrid anapole also displays much better confined internal fields. **Figure 2**b demonstrates the average electromagnetic energy density inside the cylinder at the hybrid anapole wavelength to exceed nine times the value of free space almost without leakage of the field outside the particle volume. On its turn, the latter further reduces the interaction with the surrounding (see the following sections).

**Intrinsic modal content of the hybrid anapole**

While Cartesian multipoles are suitable for the description of far-fields, in this section we employ the natural QNM expansion[52] of near-fields and internal currents , which in the following will allow us to further unveil the physics behind the hybrid anapole state. QNMs provide a suitable basis for the induced polarization currents, which takes the following form:

$$\mathbf{J}(\omega,\mathbf{r}) = \sum_s \alpha_s(\omega)\tilde{\mathbf{J}}_s(\omega,\mathbf{r}) - i\omega\delta\varepsilon\mathbf{E}_{inc}(\omega,\mathbf{r}). \qquad (2)$$

Here $\tilde{\mathbf{J}}_s(\mathbf{r}) = -i\omega\delta\varepsilon\tilde{\mathbf{E}}_s(\mathbf{r})$, $\alpha_s(\omega)$ and $\delta\varepsilon$ are, respectively, the induced modal scattering current distribution as a function of the internal mode field, the excitation coefficient of the *sth* mode describing its contribution to the total current at a given frequency, and the permittivity contrast with the host environment.

We use a modified version of the freeware MAN developed by the authors of Ref.[51]. More details on the approach can be found in the Supporting Information. For simplicity, we consider a dispersionless, lossless nanocylinder with a constant refractive index $n \approx 3.87$ (corresponding to aSi at 780 nm), so that the excitation coefficients depend solely on the fields of an individual QNM[51]. Losses and refractive index dispersion of the original design are negligible in the considered spectral range (see Supporting Information), and therefore this approximation does not change significantly the scattering cross section and average electromagnetic energy density.

The results of the QNM expansion are displayed in the different panels of **Figure 2**. The correctness of our calculations in the studied spectral range, particularly near the scattering dip, is well validated in **Figures 2**a-b by comparing the sum of the individual QNM contributions with the numerically obtained total scattering cross section (a) and average electromagnetic energy density inside the cylinder (b). From hereon, we shall label the QNMs with the standardized notation for the modes of isolated cylindrical cavities[55], i.e. $(TE, TM)_{uv\ell}$, where the sub-indices denote the number of standing wave maxima in the azimuthal $(u)$, radial $(v)$ and axial $(\ell)$ directions. *TE* and *TM* indicate the predominant nature of the internal field distribution. Specifically, *TM* (transverse magnetic) modes have $H_z \approx 0$, while *TE* (transverse electric) have $E_z \approx 0$.

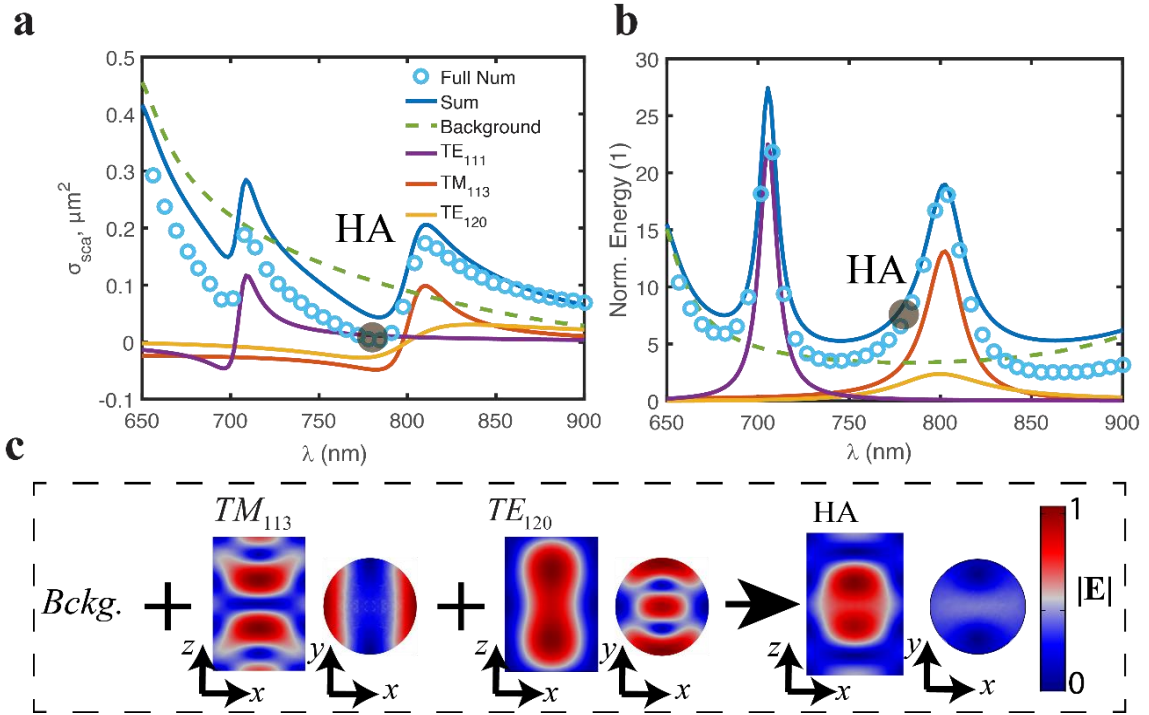

**Figure 2.** (a) Alternative scattering cross section decomposition by means of the QNM expansion method. The full-wave simulation is nearly the same (without losses), as in **Figure 2**, but in a linear scale. Colored lines are the individual contributions of the physical QNMs. The contributions of modes having their resonances outside the considered spectral range are added up in the green dashed line. Resonant modes in the considered spectral range are associated with the $TE_{111}$, $TE_{120}$ and $TM_{113}$ modes of the isolated cylinder. The blue line corresponds to the reconstructed scattering cross section, confirming that all the resonant spectral features can be successfully recovered via this method, and demonstrating good agreement near the hybrid anapole, point HA. (b) Spectra of the volume-averaged electromagnetic field energy inside the cylinder, and individual contributions of the excited modes. Colors and legends are the same as in (a). The electromagnetic energy density has been normalized with respect to the incident electromagnetic energy density in the vacuum $w_{EM} = \varepsilon_0 E_0^2$. Excellent agreement is obtained with the full-wave simulations. (c) Normalized internal electric field distributions of the two most relevant modal contributions near point A, from left to right, associated with Fabry-Perot ($TM_{113}$) and Mie-like ($TE_{120}$) standing wave patterns ($TE_{111}$ is very weak near the hybrid anapole), respectively. Their addition via Eq. (2), together with the background modes (Bckg.), leads to the reconstruction of the internal fields of the hybrid anapole, also displayed on the right-hand side of (c). All the electric fields have been normalized with their respective maxima, to enhance their visualization.

The spectral behavior of each resonant QNM is described by a Fano lineshape[56], (see Supporting Information, section 5). The other nearby QNMs constitute the background scattering contribution of the particle.

In **Figure 2**a we note that a total of three QNMs resonate in the visible range. The correct reconstruction of the scattering cross section requires taking into account background modes, despite their resonances being outside the considered spectral range (green dashed line). Nevertheless, at point HA, only the $TM_{113}$ and $TE_{120}$ modes present a 'Fano-like' response.

Now the pronounced low scattering regime can be easily grasped as a consequence of modal interference: a clear sign that this is indeed the case are the resonant negative contributions to scattering presented by both the $TE_{120}$ and $TM_{113}$ modes. It implies that, when the incident field impinges in the resonator, energy exchange takes place between the two and the background QNM fields[57]. This owes to the fact that the QNMs do not obey the usual conjugate inner product relation of orthogonal modes in Hermitian systems[58]. Here it is important to emphasize the unusual feature of the hybrid anapole: the two resonant QNMs dominating the spectra are *simultaneously negatively suppressed* by interference with the background. For comparison, the QNM decomposition of a conventional dipole anapole disk is given in the Supporting Information.

A completely different picture arises within the resonator. **Figure 2**b presents the modal decomposition of the internal energy stored in the cylinder in the vicinity of point HA. This is one of the main results of the section, since, contrarily to the multipole expansion, the QNM decomposition allows us to clearly distinguish the contributions of the eigenmodes to the internal fields. Firstly, we note that electromagnetic energy is significantly enhanced (around nine times) with respect to the incident plane wave. Secondly, it is clearly seen that the stored energy at the hybrid anapole is mainly driven by the $TM_{113}$ mode due to its higher quality factor and the proximity of its resonant wavelength to the hybrid anapole wavelength, in a minor measure by the $TE_{120}$ and the sum of the background contributions. Energy exchange between the internal fields of the QNMs is strongly minimized, as reflected in the fact that no negative contributions to the internal energy can be appreciated. Overall, the QNM analysis given in **Figure 2** demonstrates that both the invisibility effect (outside the cylinder) and the internal energy enhancement at the hybrid anapole state are mediated by the simultaneous resonant response of the $TM_{113}$ and the $TE_{120}$ modes. The background modes, on the other hand, while they do not apparently define the spectral features of the figures of merit significantly, also play an important role since their interference with the resonant ones gives rise to the invisibility effect. This interpretation is consistent with early investigations regarding the formation of Fano lineshapes in the scattering cross section of spherical resonators[53].

The electric field distributions of the $TE_{120}$ and $TM_{113}$ modes are shown in **Figure 2**c. Following Refs.[59,60] we can classify the first as a 'Mie' type mode, similar to the ones supported by an infinite cylinder, while the second is of the 'Fabry-Perot' (FP) type [59], arising due to the formation of a standing wave pattern between the top and bottom walls of the resonator, i.e. having non-zero axial wavenumber ($\ell \geq 1$). Their distinct origin unveils the reason why it is possible to obtain a hybrid anapole state in this particular geometry, contrarily to spherical scatterers. As shown in the Supporting Information, the real parts of the eigenwavelengths of the modes in the cylinder can be estimated as[61]

$$\lambda_{uv\ell} \approx \frac{\pi D}{n_p \sqrt{\left(\frac{\ell \pi}{2}\frac{D}{H}\right)^2 + z_{uv}^2}}, \qquad (3)$$

where $z_{uv}$ is the $v$th root of the $u$th Bessel function of the first kind for *TE* modes, or its first derivative for *TM* modes. For FP modes, $\ell \neq 0$ and the denominator in Eq. (3) displays a strong dependence on the aspect ratio $D/H$ of the cylinder. In contrast, since $\ell = 0$ for Mie modes, their eigenwavelengths only change with $D$. Thus, the eigenwavelengths and the multipolar content of these two mode types *are independently tunable* from each other, resulting in a flexible control over the optical response of the resonator and enabling the simultaneous scattering suppression observed in **Figure 1**a-b, and **Figure 2**a - the hybrid anapole state.

A straightforward comparison between the QNM and multipolar methods allows determining the multipoles radiated by a given QNM (see section 10, Figure S4, in the Supporting Information). Specifically, Figure S3 leads us to the conclusion that the $TE_{120}$ mode radiates primarily as $p$, with a minor $Q^{(m)}$, while the $TM_{113}$ mode scatters as a combination of $m$ and $Q^{(e)}$ (in both cases, referring to both their basic and toroidal counterparts). When scattering from a mode is resonantly suppressed, radiation from the multipoles associated to it is also strongly minimized, and results in the different multipole anapole states observed in the decomposed spectra of **Figure 2**a. Thus, the close proximity of the destructive interference points of the $TE_{120}$ and $TM_{113}$ modes at point HA leads to an overlap of the anapole states of the four dominant multipoles. In this fashion, using the QNM expansion approach, we have shown an alternative and general physical explanation of dark scattering states and qualitatively illustrated its link to the multipolar response of the particle. This self-consistent approach will further be effectively applied to understand the physics behind the particle-substrate interaction.

As a final remark, we point out that the quality factors of the $TE_{120}$ and $TM_{113}$ modes in the structure are, respectively, 12 and 33. A direct comparison with the $TE_{120}$ supporting a conventional anapole disk (see Supporting Information) shows that the quality factor of the $TM_{113}$ mode at the hybrid anapole is more than four times larger. Therefore, we anticipate much better performance of the hybrid anapole for *second and third harmonic generation* processes with respect to conventional anapole disks, since nonlinear scattering cross sections scale linearly with the quality factors of the modes involved[62].

**Substrate-mediated transient signal modulation**

We will now proceed to illustrate in detail a novel effect, unique to the hybrid anapole, by which its nonradiating nature can be preserved in the stationary regime when deposited on any dielectric substrate. We then exploit the underlying mechanism behind the phenomena to modulate the transient response of the isolated nanocylinder.

In comparison with any other scattering phenomena including conventional anapoles, the hybrid anapole state is remarkably robust in the presence of a substrate. This particular behavior is illustrated in **Figure 3**a, where we have plotted the calculated scattering cross sections of the cylinder in free space and deposited over glass ($n_s = 1.5$), hypothetical substrates with $n_s = 2, 3$ and over amorphous silicon ($n_s = 3.87$). The amplitude and spectral position of the scattering dip are almost undisplaced from the free-space values while the refractive index contrast at the bottom of the particle remains non-zero. However, we observe drastic changes in the shape of the Fano profile, once the contrast is absent (silicon particle over silicon substrate). There exists a nontrivial underlying mechanism by which the hybrid anapole is "protected" against modifications of the substrate refractive index. The physics becomes transparent when examining the distinct nature of the two resonant modes involved in the formation of the hybrid anapole (**Figure 3**c). On the one hand, increasing the refractive index of the substrate leads to a decrease in the lower wall reflectivity, which is crucial for the standing Fabry-Perot mode $TM_{113}$ inside the resonator. Approaching $n_s$ to the particle's refractive index results in a higher energy leak towards the substrate and a substantial decrease in the quality factor $Q_\ell$. Consequently, the resonance flattens and disappears when the lower boundary becomes transparent, as shown in **Figure 3**b.

The standing wave pattern of the $TM_{113}$ resembles a one-dimensional Fabry-Perot cavity on a dielectric substrate. The QNMs of this simplified model has the advantage of being analytically solvable[52], thus providing valuable physical insight easy to extrapolate to the problem at hand. As we derive in the Supporting Information, the QNMs are formed by two interfering plane waves traveling in opposite directions inside the cavity, when the driving wavelength satisfies the condition

$$r_{21} r_{23} w^2 = 1, \qquad (4)$$

where $w = \exp(ik_\ell n_r H)$ and $r_{21}$, $r_{23}$ are the Fresnel reflection coefficients from the cavity-air and the cavity-substrate interfaces, respectively. The quality factor of a QNM with axial index $\ell$ is calculated as

$$Q_\ell = -\frac{\ell \pi}{\ln(r_{23} r_{21})}. \qquad (5)$$

Equation (5) serves very well to illustrate the influence of the substrate on the $TM_{113}$ mode. When the two reflection coefficients are unity, the energy is completely stored inside the resonator and $Q_\ell$ is infinite. Similarly, a decrease in the reflection coefficient from the substrate leads to radiative losses and a decrease in the quality factor, effectively becoming zero when the contrast is absent. Indeed, a lower quality factor leads to less appreciable spectral features, as observed in the simulations (**Figure 3**b). Another important conclusion that one can draw from the numerator in Eq. (5) is that the real part of the resonant wavelength of the QNM is independent of the refractive index at the walls. Thus, modifying the substrate refractive index does not shift the spectral position of the resonance (i.e. does not shift the hybrid anapole wavelength), but simply changes the amplitude and width of the Fano profile.

With proper normalization, and employing the notation of the inset in **Figure 3**b, the amplitudes of the incoming and outgoing plane waves inside the resonator are $\left|A_2^+\right| = \sqrt{\alpha r_{12} w}$ and $\left|A_2^-\right| = \sqrt{\alpha r_{23} w}$, with $\alpha = 1/(4H\varepsilon)$ (see Supporting Information for details). Plane waves reflected from the substrate are thus decreased with decreasing index contrast. Consequently, the same occurs to the lower field maxima of the $TM_{113}$. While this prediction is observed in the simulations for relatively low contrasts, the behavior at very small contrasts is very different (see the case with $n_s = 3$ in **Figure 3**c). In the latter situation, the standing waves along z become negligible in comparison with the initially weaker standing waves in the x-y plane, and the QNM can only be well described numerically.

Contrarily, it is noteworthy that even comparably small contrast (3 – 3.87) leads to enough contribution of the $TE_{120}$ mode to still preserve the scattering dip (see **Figure 3**b). The standing wave pattern of this second mode develops in the lateral walls of the cylinder, and therefore depends much less on variations of the reflectivity of the lower wall, keeping an almost constant quality factor (see Supporting information). Most of the QNM energy is then stored in the resonator even in the case of zero effective contrast with the substrate, as demonstrated in **Figure 3**c. It results in a larger contribution of the $TE_{120}$ mode to extinction at small contrasts, and a decrease of radiation from the electric quadrupole and magnetic dipole. The hybrid anapole is now mainly driven by electric dipole radiation stemming from the $TE_{120}$ mode (see Supporting Information). Thus, for large $n_s$ the hybrid anapole *degenerates into a conventional electric dipole anapole state*, but still retains its non-radiative nature, since the other multipoles become negligible due to the small reflections at the lower wall of the nanocylinder.

While the effect might not seem remarkable at first glance, it opens the possibility to realize *ultrafast time modulation of the scattered signal* in the transient regime. To illustrate this novel concept, we outline here a temporal coupled mode (TCM) scheme to model the transient behavior of the resonator with a set of linear equations:

$$\frac{d}{dt} a_i = \left(-i\omega_i - \gamma_i\right) a_i + \sqrt{\gamma_i} s_i^+ \tag{6}$$

$$s_i^- = s_i^+ e^{i\phi} + \sqrt{\gamma_i} a_i, \tag{7}$$

here $s_i^{+(-)}$ are the input (output) amplitudes of a given scattering channel coupled to a specific mode, and the $a_i$ coefficients correspond to the amplitudes of the excited modes $TM_{113}$ and $TE_{120}$, normalized to their stored energies inside the resonator, with resonant frequencies $\omega_i$ and radiative losses $\gamma_i$. The phase factor $e^{i\phi}$ emulates the contribution of the background modes. Eqs. (6)-(7) take into account that no coupling can occur between the resonant modes involved, (the derivation can be found in the Supporting Information).

In the absence of incident radiation (all $s_i^+ = 0$), the solution of Eq. (6) for each of the two modes shows that their field amplitudes $E_i \propto a_i$ follow a time dependence of the form $e^{-i\omega_i t} e^{-\gamma_i t}$. Due to the mismatch between their resonant frequencies, and the fact that their radiative losses have the same order of magnitude, the total scattered field experiences beatings with a low frequency envelope given by $\cos((\omega_1 - \omega_2)/2)$, modulating a high frequency signal $\cos((\omega_1 + \omega_2)/2)$, where $\omega_1$ and $\omega_2$ are the resonant frequencies of the $TM_{113}$ and $TE_{120}$ QNMs, respectively.

Tuning the substrate refractive index allows modifying the stored energy of the $TM_{113}$ mode, effectively controlling the beating amplitudes in the transient.

In the simulations shown in **Figure 3**d-g, the hybrid anapole nanoparticle is excited by a plane wave square pulse with duration $\tau = 200\,\text{fs}$, (see inset in **Figure 3**d) sufficiently long to allow the observation of both stationary and transient regimes. The scattered power is then plotted in **Figure 3**d, and a more detailed view of the transient after the pulse is shown in **Figure 3**e for the different refractive index contrasts. Clear differences between the beating patterns can be appreciated, enabled by the leakage of the Fabry-Perot modes discussed above. In all cases, however, the decay of the Mie mode $TE_{120}$ is perfectly visible since it is barely affected by the substrate. The retarded peak that gradually disappears with decreasing contrast is associated with the slow decay of the Fabry-Perot mode $TM_{113}$.

This behavior is reflected on the recorded spatiotemporal maps of the $E_x$ field in **Figures 3**f-g,

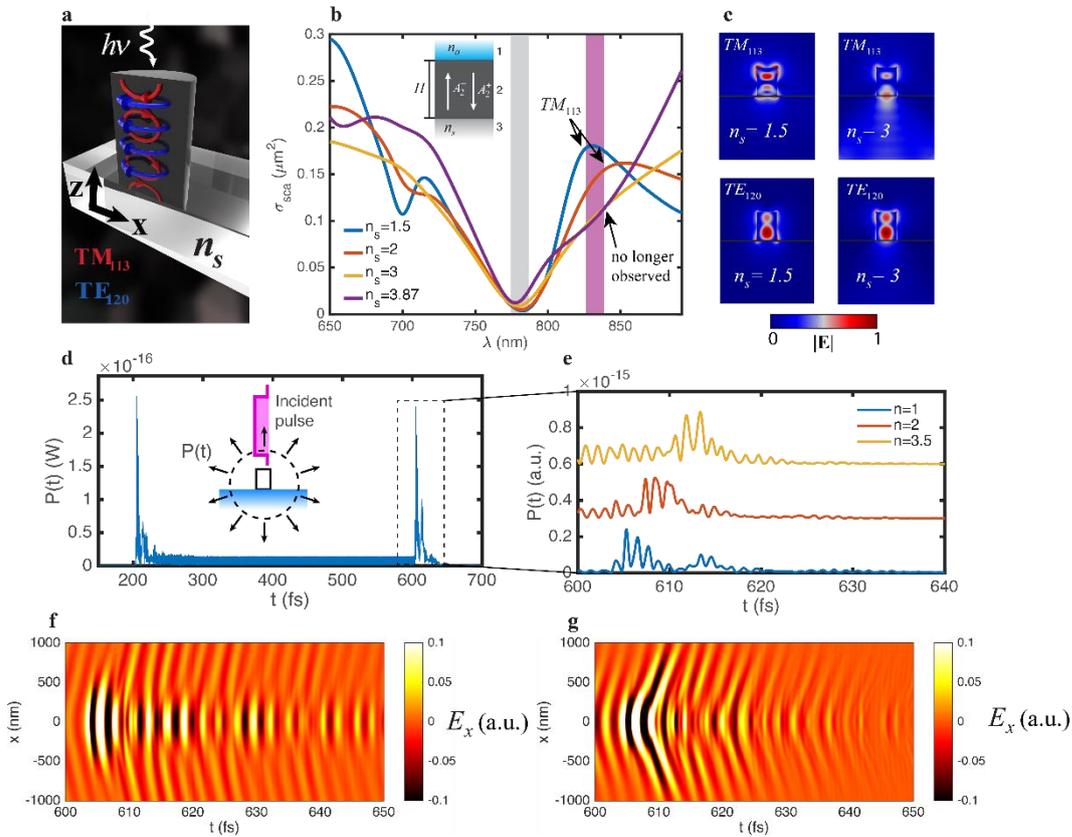

**Figure 3**. Stationary and transient response of the hybrid anapole with different index contrasts with the substrate. (a) Scheme depicting the setup and the excited eigenmodes. (b) Comparison between the numerically obtained scattering cross sections for the nanocylinder with the size from **Figure 1**, deposited on substrates with increasing refractive index. The calculations are performed with the full experimental aSi refractive index given in the Supporting Information. Inset of (b): One-dimensional Fabry Perot model of the $TM_{113}$ mode. (c) xz field distributions of the QNMs $TE_{120}$ and $TM_{113}$ when the cylinder is deposited over substrates with different refractive index. As predicted by our theory, losses from the $TM_{113}$ mode increase when decreasing the refractive index contrast. Contrarily, the $TE_{120}$ mode remains confined in the scatterer. (d-f): Temporal response of the hybrid anapole under a plane wave square pulse; (d) Scattered power as a function of time; (e) Different beating patterns of the decaying eigenmodes after the plane wave excitation as a function of $n_s$; (f-g) Spatiotemporal maps of the x-component of the scattered electric field

measured along the x axis for $n_s = 1$ (f) and $n_s = 2$ (g), measured from the beginning of the modal decay. The colorbars are saturated for better visualization.

where the characteristics of the mode beating of the system can be fully grasped. At high contrasts, the $TM_{113}$ mode leaks from the resonator, and a well-defined beating pattern is observed. The beating is even along the x-axis, reflecting the symmetry of the fields of the dominant modes. At lower contrast, the decreased contribution of the $TM_{113}$ mode leads to damping at longer times. This is explained by the smaller Q-factor of the $TE_{120}$ mode, which results in a shorter transient.

**Non-Huygens transparent metasurface with controllable transients**

The potential that transient modulation offers can be fully exploited in practice by fabricating metasurfaces inheriting (and enhancing), the single particle mechanism described in the previous section.

In particular, the effect can be harnessed to design fully transmissive, all-dielectric metasurfaces without relying on the well-known Huygens condition[6,63], and simultaneously realize controllable ultrafast switching by harnessing transients (see **Figure 4**a-d). Contrarily to the latter, the light traverses the structure without significant phase variation, thus rendering the metasurface invisible (see shadowed selection in **Figure 6**b). This is a direct consequence of Eq.(1). It can be easily seen by writing the transmission coefficient as a sum of the relevant multipole contributions of the meta-atoms[64]:

$$t = 1 + t_p + t_m + t_{Q^{(m)}} + t_{Q^{(e)}} \tag{8}$$

Each term in the previous sum is proportional to the corresponding total multipole moment (basic and toroidal contributions).

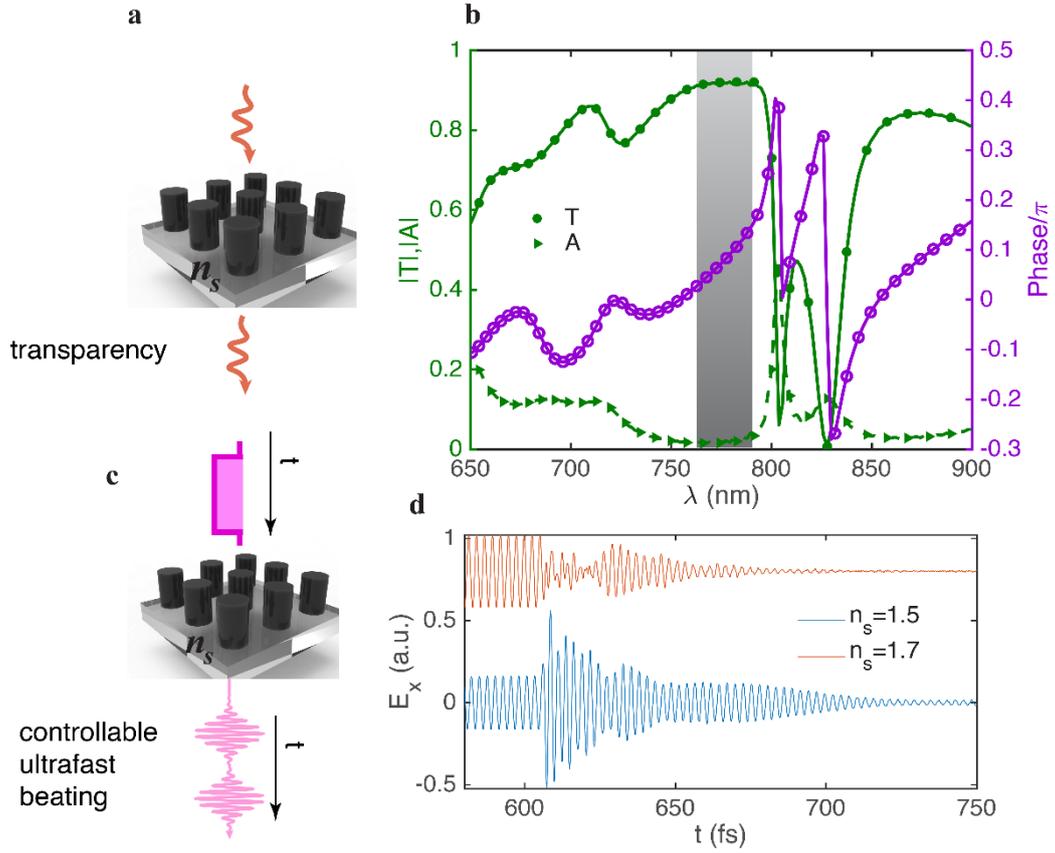

**Figure 4.** Design of a subwavelength hybrid anapole-based metasurface, featuring double functionality in the stationary and ultrafast regimes. The period is set to 530 nm. (a) artistic representation of the proposed metasurface operating at the hybrid anapole state in the stationary regime. The structure is illuminated by an x-polarized plane wave, passing through the array completely undistorted. (b) Transmission, absorption and phase of the transmitted wave with respect to the incident one. As predicted by Eqs. (1) and (8), the out-of-phase basic and toroidal moments induce a transparency band with very small phase perturbation (shaded region). The calculations have been performed in the presence of a substrate with $n_s = 1.5$. (c) Artistic representation illustrating the metasurface response in the sub-ps regime, when excited by a plane wave pulse. (d) $E_x$ component of the scattered field from the metasurface as a function of time at the end of the plane wave pulse, for $n_s = 1.5$ and after increasing the substrate index by 0.2. The results demonstrate that transient oscillations of the metasurface are highly sensitive to changes in the substrate index, opening a pathway towards tunable ultrafast beating.

When Eq. (1) is fulfilled, the multipolar contributions in Eq. (8) are zero. Consequently, we are left with $t \approx 1$ in Eq. (8), and the incident wave leaves the system unperturbed (see **Figure 4**a). This condition is closely fulfilled in a transmission band around the hybrid anapole wavelength ($\lambda = 782$ nm as previously), as indicated in **Figure 4**b. Polarization losses induce absorption, which slightly decreases transmission. The considered period (530 nm) is not unique, i.e. once the geometry supporting the hybrid anapole for an isolated particle is known, a subwavelength metasurface of such particles will mimic the single particle behavior far away from the first diffraction order[65].

In analogy with the single particle behavior, the metasurface induces ultrafast beating of the scattered field in the transients at the beginning and the end of a plane wave pulse (see **Figures 4**c-d). Remarkably, however, the sensitivity to the contrast with the underlying substrate is significantly accentuated, inducing large changes in the modulated scattered signal with an increase of $n_s$ of only 0.2, as can also be observed in the supplementary video, where the complete picture on the evolution of the near field topology as a function of time is presented.

The sub-ps relaxation times of the metasurface, together with the high substrate sensitivity of the transients, opens an exciting pathway towards the spatiotemporal control of ultrafast dynamic effects. Among the interesting perspectives, ultrashort laser pulses can be effectively modulated in time by tailoring the transient phenomena. The high substrate sensitivity can be exploited in order to dynamically tune the transient response of the metasurface in different temporal regimes by selecting the appropriate modulation technology[66]. The bulk refractive index of the substrate can be tuned, for example, by means of phase-change materials such a GeTe[67] using temperature as a control parameter. More sophisticated, yet much faster refractive index changes could be achieved by tuning optical nonlinearities arising in the substrate as a function of the incident beam intensity (e.g. if ITO constitutes the underlying substrate[68]). Finally, modifying the free electron density in the substrate with electrical gating[69] might constitute a promising approach to achieve the desired tunability in practical applications.

**Discussion**

In the stationary regime, we have theoretically predicted and experimentally confirmed the existence of hybrid anapoles, exotic non-scattering states requiring the simultaneous destructive interference of electric and magnetic cartesian multipoles with their toroidal counterparts. For the first time, dynamic toroidal quadrupoles, as well as magnetic anapoles have been confirmed experimentally in isolated subwavelength nanoparticles. In order to broaden our knowledge of anapole electrodynamics beyond the electric dipole approximation, we have developed a solid physical picture[16] in terms of the fundamental eigenmodes driving the resonator response. Magnetic anapoles and quadrupole anapoles display field confinements several times larger than electric anapoles, promising higher enhancements of nonlinearities in the absence of linear scattering.

In the transient regime, the prior analysis has allowed us to design at will the breakdown of the hybrid anapole conditions to obtain ultrafast modulation of the scattered power. The present theory shows that efficient spatiotemporal control of the transients mediated by the underlying substrate can be achieved, while maintaining scattering in the stationary regime at vanishing level. Following the established theoretical guidelines, we propose and verify numerically a non-Huygens, subwavelength metasurface obeying the same principles and therefore featuring a double functionality: (i) substrate-independent full transparency in both amplitude and phase mediated by the hybrid anapole in the stationary regime, and (ii) ultrafast, substrate-dependent signal modulation in the transient regime, mediated by its breakdown. The necessary changes in the substrate refractive index to observe appreciable modifications of the transient response of the metasurface are found to be well in the range of available technologies for dynamic tuning[67], drastically expanding the available degrees of freedom.

Therefore, combining our findings with up-to-date modulation techniques holds great promise for future applications in the emerging field of ultrafast dynamic nanophotonics. Particularly, the ultrashort scattering peaks arising at the transient can be used to enhance ultrafast nonlinearities[70] (e.g. single particle lasing), or coded in terms of their intensity as a function of the substrate index to realize light-based computing[71]. We note from the presented theory that transients oscillate at frequencies shifted from the original driving pulse, and therefore offer the possibility to realize linear frequency conversion[72]. Finally, the proposed metasurface can enable the temporal/spectral shaping of fs laser pulses by interfering the latter with the out-going scattered signal.